\documentclass[12pt,a4paper]{article}
\usepackage{a4wide}
\usepackage{amsmath}
\usepackage{amssymb}
\usepackage{amsfonts}
\usepackage{epsfig}
\usepackage{subfigure}
\usepackage{exscale}
\usepackage{float}
\usepackage{bbm}
\usepackage[numbers,sort&compress]{natbib}

\newcommand{\p}{\partial}

\makeatletter
\@addtoreset{equation}{section}
\makeatother

\title{Constraint Effective Potential of the Magnetization in the Quantum
XY Model}

\author{U.\ Gerber$^a$, C.\ P.\ Hofmann$^b$, F.-J.\ Jiang$^c$, G.\ Palma$^d$,
P.\ Stebler$^a$, \\ and U.-J.\ Wiese$^a$
\\ \\
$^a$ Albert Einstein Center for Fundamental Physics \\
Institute for Theoretical Physics, Bern University \\
Sidlerstrasse 5, CH-3012 Bern, Switzerland
\\ \\
$^b$ Facultad de Ciencias, Universidad de Colima \\
Bernal D\'iaz del Castillo 340, Colima C.P.\ 28045, Mexico
\\ \\
$^c$ Department of Physics, National Taiwan Normal University \\
88, Sec.4, Ting-Chou Rd., Taipei 116, Taiwan
\\ \\
$^d$ Departamento de F\'isica, Universidad de Santiago de Chile \\
Casilla 307, Santiago 2, Chile \\ \\}

\begin{document} 

\bibliographystyle{plain}

\maketitle

\begin{abstract} \normalsize

Using an improved estimator in the loop-cluster algorithm, we investigate the 
constraint effective potential of the magnetization in the spin $\tfrac{1}{2}$ 
quantum XY model. The numerical results are in excellent agreement with the 
predictions of the corresponding low-energy effective field theory. After its
low-energy parameters have been determined with better than permille precision,
the effective theory makes accurate predictions for the constraint effective
potential which are in excellent agreement with the Monte Carlo data. This shows
that the effective theory indeed describes the physics in the low-energy regime
quantitatively correctly.

\end{abstract}

\newpage

\section{Introduction}

When a nonperturbative system of many strongly coupled degrees of freedom 
undergoes the spontaneous breakdown of a continuous global symmetry, massless
Goldstone bosons arise as the relevant low-energy degrees of freedom. Despite
the fact that the underlying microscopic system can usually only be studied
numerically, the low-energy dynamics of the Goldstone bosons can be addressed
analytically using a systematic low-energy effective field theory. The a priori
unknown low-energy parameters of the effective theory can then be determined by
matching the results of numerical simulations of the underlying microscopic 
system to analytic results of the effective field theory. Thanks to this 
interplay between numerical simulations and analytic effective field theory
calculations, important insights have been gained into the Higgs sector of the
Standard model \cite{Goe91b,Goe93} and the dynamics of QCD as well as of 
magnetic systems, including the undoped precursors of high-temperature 
superconductors \cite{Wie94,Bea96}. The latter are described by a low-energy 
effective theory for magnons \cite{Cha89,Neu89,Fis89,Has91,Has93} --- the 
Goldstone bosons of the spontaneously broken $SU(2)$ spin symmetry. Using 
this theory, the shape of the constraint effective potential of the order 
parameter has been worked out in detail by G\"ockeler and Leutwyler 
\cite{Goe91,Goe91a}. Their predictions for the shape of the constraint 
effective potential had already been tested against Monte Carlo simulations
of classical 3-d $O(3)$ and 4-d $O(4)$ lattice models \cite{Dim91}. Recently,
we have performed a high-accuracy investigation of the antiferromagnetic spin 
$\tfrac{1}{2}$ quantum Heisenberg model on a square lattice \cite{Ger09}. In 
particular, we have employed a new improved estimator for the distribution of
the staggered magnetization using the loop-cluster algorithm
\cite{Eve93,Wie94,Bea96}. The very accurate Monte Carlo data were then
compared with the analytic results of the magnon effective theory, resulting
in a determination of the low-energy parameters with permille accuracy. In
this paper, we extend this investigation to the $(2+1)$-d quantum XY model,
which models quantum magnetism, as well as superfluidity of hard-core bosons and has a spontaneously broken
$U(1)$ symmetry. The analytic results of the effective field theory readily
apply to the Goldstone bosons of superfluidity. In the present paper, however, we will use ``magnetic'' language throughout, having in mind quantum magnetism. After its low-energy parameters
have been determined with better than permille precision, the effective theory
makes unambiguous predictions for the constraint effective potential, which are
in excellent agreement with the Monte Carlo data.

The rest of this paper is organized as follows. In section 2 we summarize the 
predictions of the low-energy effective field theory. In section 3 we present 
the results of our numerical simulations obtained with the loop-cluster algorithm, which
are compared with the effective field theory predictions in section 4. Finally, 
section 5 contains our conclusions.
 
\section{Effective Field Theory Predictions}

In section 4 we will compare our very accurate Monte Carlo data with the effective 
field theory predictions of G\"ockeler and Leutwyler \cite{Goe91,Goe91a} which 
are derived from a scalar 3-d $O(N)$-symmetric effective field theory. Here, we 
summarize those results that are relevant for our study. While these results
were derived in the framework of a 3-d relativistic quantum field theory, they
are readily applicable to the $(2+1)$-d quantum XY model, because at 
low energies its Goldstone boson has a linear ``relativistic'' dispersion
relation. The Hamiltonian of the quantum XY model is defined as
\begin{equation}
H = - J \sum_{\langle xy \rangle} (S^1_x S^1_y + S^2_x S^2_y) - \vec M \cdot \vec B,
\end{equation}
where $x$ and $y$ are nearest-neighbor sites on a square lattice with spacing
$a$, and $J>0$ is a constant. Working in natural units in which $\hbar = 1$, the spin $\tfrac{1}{2}$
operators $\vec S_x$ obey the standard commutation relations 
\begin{equation}
[S_x^a,S_y^b] = i \delta_{xy} \varepsilon_{abc} S_x^c.
\end{equation}
The quantity $\vec B = (B_1,B_2)$ is a uniform magnetic field in the XY plane 
that couples to the magnetization order parameter
\begin{equation}
\vec M = \Big(\sum_x S^1_x,\sum_x S^2_x \Big).
\end{equation}
In the infinite volume limit and at zero temperature, the vacuum expectation
value of $\vec M$ is non-zero, signaling the spontaneous breakdown of the 
$U(1)$ spin symmetry, which is generated by the 3-component of the total spin
\begin{equation}
S^3 = \sum_x S^3_x.
\end{equation}
It should be noted that, in contrast to the Heisenberg model which has an 
$SU(2)$ spin symmetry, even at $\vec B = 0$, here only the generator $S^3$
commutes with the Hamiltonian. It should also be pointed out that, on a bipartite
lattice, in the absence of a magnetic field, both the antiferromagnetic and the ferromagnetic XY model describe the 
same physics, since they are related to each other by a unitary transformation. 

At low energies, the relevant degrees of freedom are the Goldstone bosons
resulting from the spontaneously broken global symmetry. In the present case of 
the XY model, the $U(1)$ spin symmetry breaks completely, and we thus have 
one magnon (or superfluid Goldstone boson, depending on the physical interpretation of the quantum XY model). The low-energy effective field theory is formulated in terms of the
magnetization order parameter field
\begin{equation}
\vec e(x) = (e_1(x),e_2(x)) \in S^1, \quad \vec e(x)^2 = 1,
\end{equation}
where $x = (x_1,x_2,t)$ is a point in Euclidean space-time. Up to higher-order
corrections in a derivative expansion, the effective action takes the form
\begin{equation}
\label{action}
S[\vec e] = \int d^2x \ dt \ \left[\frac{\rho}{2}
\left(\p_i \vec e \cdot \p_i \vec e + 
\frac{1}{c^2} \p_t \vec e \cdot \p_t \vec e \right)
- {\cal M} {\vec e} \cdot {\vec B} \right],
\end{equation}
where $\rho$ is the spin stiffness, $c$ is the spin-wave velocity, and
$\cal M$ is the magnetization density. The partition function is then given by
\begin{equation}
Z = \int D\vec e \ \exp(- S[\vec e]).
\end{equation}
The dispersion relation of the magnon takes a ``relativistic'' form, where the 
velocity of light is replaced by the spin-wave velocity $c$. By introducing 
$x_3 = c t$, the effective action can thus be written in the manifestly 
Euclidean space-time rotation-invariant form
\begin{equation}
S[\vec e] = \int d^3x \ \frac{1}{c} \left( \frac{\rho}{2}
\p_\mu \vec e \cdot \p_\mu \vec e - {\cal M} {\vec e} \cdot {\vec B} \right).
\end{equation}
Since the underlying XY model does not exhibit Euclidean rotational invariance,
the symmetry is accidental and exists only in the leading terms of the 
derivative expansion. Indeed, higher-order four-derivative terms in the 
effective action break Euclidean space-time rotation invariance down to the 
discrete rotation subgroup of the square lattice.

The system, described by the effective action (\ref{action}), is considered in a
periodic cubic space-time volume $L \times L \times \beta$ with the inverse
temperature fixed at $\beta = L/c$.\footnote{We set the Boltzmann constant 
$k_B=1$.} The space-time average of the magnetization is given by
\begin{equation}
\vec \Phi = \frac{1}{2} \frac{1}{L^2 \beta} \int d^2x \ dt \ \vec e(x) =
\frac{1}{2} \frac{1}{L^3} \int d^3x \ \vec e(x).
\end{equation}
In contrast to \cite{Goe91,Goe91a}, we have included a factor $\tfrac{1}{2}$ in 
the definition of $\vec \Phi$ because the quantum spins of the underlying XY 
model have $S = \tfrac{1}{2}$, while the effective field $\vec e(x)$ is 
normalized to 1. The probability distribution of the mean magnetization 
$\vec \Phi$ is obtained as a $\delta$-function constrained path integral for 
the partition function
\begin{equation}
p(\Phi) = \frac{1}{Z} \int D\vec e \ \exp(- S[\vec e]) \ 
\delta\left(\vec \Phi - \frac{1}{2} \frac{1}{L^3} \int d^3x \ \vec e(x)\right).
\end{equation}
As a consequence of the $U(1)$ symmetry, it only depends on the magnitude 
$\Phi = |\vec \Phi|$. This distribution is normalized according to
\begin{equation}
\label{norm}
2 \pi \int_0^\infty d\Phi \ \Phi \ p(\Phi) = 1.
\end{equation}
One of the fundamental quantities in the present study is the constraint 
effective potential $u(\Phi)$. It represents the free energy density of the 
model which is obtained by computing the path integral over configurations 
constrained to a given fixed mean magnetization value $\Phi$, and is determined 
by
\begin{equation}
p(\Phi) = {\cal N} \exp(- L^3 u(\Phi)).
\end{equation}
The analytical expression for the normalization factor ${\cal N}$, derived in 
\cite{Goe91} reads
\begin{equation}
\label{normalization}
{\cal N} = \frac{1}{\widetilde{\cal M}^2} \ 
\frac{\rho L}{4 \pi^2 c e^{\beta_0/2}} \ 
\left[1 + {\cal O}\left(\frac{1}{L^2}\right) \right],
\end{equation}
where we have defined the magnetization per spin
$\widetilde{\cal M} = {\cal M} a^2$. The quantity $\beta_0$ is a 
shape-dependent coefficient characterizing the geometry of the space-time box. 
For the exactly cubical space-time volume considered here it is given by
$\beta_0 = 1.45385$. In the infinite-volume and zero-temperature limit the 
constraint effective potential approaches the infinite volume effective 
potential which is known to be a convex function of $\Phi$ \cite{Rai86,Mac05}. 
In a finite volume, on the other hand, $u(\Phi)$ is not necessarily convex. We 
may define an extensive variant of the intensive quantity $u(\Phi)$ as
\begin{equation}
U(\Phi) = L^3 u(\Phi).
\end{equation}

Within the effective field theory framework, the finite-size corrections to
the constraint effective potential were systematically worked out by
G\"ockeler and Leutwyler \cite{Goe91,Goe91a}. Near its minimum the $1/L$ 
expansion of $U(\Phi)$ takes the form
\begin{equation}
\label{UPhi}
U(\Phi) = U_0(\psi) +  \frac{c}{\rho L} \ U_1(\psi)
+{\cal O}\left(\frac{1}{L^2}\right).
\end{equation}
The quantities $U_0(\psi)$ and $U_1(\psi)$ depend on $L$ only through the rescaled variable
\begin{equation}
\label{psi}
\psi = \frac{\rho L}{c} \ \frac{\Phi - \widetilde{\cal M}}
{\widetilde{\cal M}}.
\end{equation}
The leading order contribution to the constraint effective potential is given 
by the inverse Laplace transform
\begin{equation}
\label{U0}
\exp(- U_0(\psi)) = \int_{-\infty}^\infty dx \ \exp(- i x \psi + \Gamma(ix))
\end{equation}
of the function
\begin{equation}
\label{g0}
\Gamma(\xi) = \frac{1}{2} \sum_{n=0}^\infty \frac{\beta_n \xi^n}{n!}. 
\end{equation}
Again, the quantities $\beta_n$ are shape-dependent coefficients of the finite
space-time box which are described in detail in appendix B of \cite{Has90}. 
Remarkably, the function $\Gamma(i x)$ is entirely kinematical and thus, unlike 
$U_1(\psi)$, the quantity $U_0(\psi)$ is universal, i.e.\ completely 
independent of the low-energy parameters \cite{Goe91a}. Hence $U_0(\psi)$ is 
the same for all 3-d systems with a spontaneously broken $U(1)$ symmetry, 
including the 3-d classical and the $(2+1)$-d quantum XY model. 
The $1/L$ correction to the leading contribution $U_0(\psi)$ is given by
\begin{eqnarray}
\label{U1}
U_{1}(\psi) = \psi + \exp(U_{0}(\psi))\int_{0}^{\infty}dx{\text{Re}}
\{\exp(-ix\psi+\Gamma(ix))\Omega(ix)\},
\end{eqnarray}
with
\begin{eqnarray}
&&\Omega(\xi) = -\frac{1}{4} \left ( \xi \omega(\xi)^2 - 2\omega(\xi) 
- \frac{\xi^2}{16 \pi^2}  \right  ) - k_0\xi^2,\nonumber\\
&&\omega(\xi)=\sum_{n=1}^{\infty}\frac{\beta_n}{(n-1)!}\xi^{n-1}
.\end{eqnarray}
Furthermore, $k_0$ appearing above is a low-energy constant which is given by
\begin{equation}
k_0 = \frac{2 \rho^3}{{\cal M}^2 c^2} (h_1 + h_2)+\frac{1}{64 \pi^2},
\end{equation}
where $h_1$ and $h_2$ are the low-energy constants associated with the higher-order terms in the effective action
\begin{equation}
\Delta S[\vec e] = - \int d^2x \ dt \
\left[h_1 (\vec e \cdot \vec B)^2 + h_2 \vec B^2\right].
\end{equation} 
A non-zero magnetic field $\vec B$ in the XY plane turns the 
magnons into pseudo-Nambu-Goldstone bosons with a non-zero mass $m$ 
determined at leading order by
\begin{equation}
m^2 = \frac{{\cal M} B}{\rho c^2}, \quad B = |\vec B|.
\end{equation}
The constant $k_0$ also appears in the $B$-dependence of the field 
expectation value
\begin{eqnarray}
\label{PhiB}
|\langle {\vec \Phi} \rangle (B)| & = & {\tilde {\cal M}} \ \Bigg\{ 1 + \frac{1}{8} { \Big( \frac{c}{\rho L} \Big)}^2
\sum_{n_1,n_2=0}^{\infty} \frac{(n_1+n_2+1) \beta_{n_1+1} \beta_{n_2+1} }{n_1! n_2!} \nonumber \\
& \times &{(mcL)}^{2n_1+2n_2} -  \frac{1}{8} { \Big( \frac{c}{\rho L} \Big)}^2 \sum_{n=0}^{\infty} \frac{2n \beta_{n+1}}{n!} {(mcL)}^{2n-2} 
\nonumber \\
& + & \frac{1}{2} \frac{c}{\rho L} \sum_{n=0}^{\infty} \frac{\beta_{n+1}}{n!} {(mcL)}^{2n}
- \frac{1}{8} {(\frac{c}{\rho L})}^2 \frac{1}{{(mcL)}^4} \nonumber \\
& - & \frac{1}{2} \frac{c}{\rho L} \frac{1}{{(mcL)}^2} - \frac{1}{64 \pi^2} {\Big( \frac{m c^2}{\rho} \Big)}^2
+ k_0 \Big( {\frac{m c^2}{\rho} \Big)}^2 + {\cal O}(m^3) \Bigg\}.
\end{eqnarray}
It should be noted that eq.(\ref{PhiB}) was derived in the $p$-regime of 
chiral perturbation theory in which $m c L \gg 1$ while $m c^2$, 
$c/L \ll 2 \pi \rho$. In particular, in eq.(\ref{PhiB}) one cannot make $B$ (and
thus $m$) arbitrarily small, because one would otherwise enter the 
$\epsilon$-regime in which $m c L \approx 1$.

The low-energy constant $k_0$ can be determined either from $U_1(\psi)$ or 
$|\langle {\vec \Phi} \rangle (B)|$ by fitting the relevant Monte Carlo data 
to the corresponding theoretical predictions (eq.(\ref{U1}) and eq.(\ref{PhiB})). As we
will demonstrate later, the numerical values for $k_0$ obtained from 
$U_1(\psi)$ and $|\langle {\vec \Phi} \rangle (B)|$ are consistent.

G\"ockeler and Leutwyler have also worked out analytic predictions for the
first and second moment of the probability distribution $p(\Phi)$ up to two 
loops. They obtained
\begin{eqnarray}
\label{moments}
&&\langle \Phi \rangle = 
\widetilde{\cal M} \left(1 + \frac{c}{\rho L} \frac{\beta_1}{2} +
\frac{c^2}{\rho^2 L^2} \frac{\beta_1^2}{8} \right)
+ {\cal O}\left(\frac{1}{L^3}\right), 
\nonumber \\
&&\langle (\Phi - \langle \Phi\rangle)^2\rangle = 
\frac{\widetilde{\cal M}^2 c^2}{\rho^2 L^2} \frac{\beta_2}{2} + 
{\cal O}\left(\frac{1}{L^3}\right),
\end{eqnarray}
where the additional shape-dependent coefficients for the cubic box considered 
here are given by $\beta_1 = 0.225785$ and $\beta_2 = 0.010608$ \cite{Has90}.

Other physical quantities of central interest are the susceptibilities. First, 
one identifies the order parameter susceptibility
\begin{equation}
\label{defstagg}
\chi_1 = \frac{1}{L^2} \int_0^\beta dt \ \frac{1}{Z} 
\mbox{Tr}[M^1(0) M^1(t) \exp(- \beta H)].
\end{equation}
Here $M^1 = \sum_x S^1_x$ is the first component of the magnetization. A second
susceptibility refers to the $U(1)$ conserved quantity $ M^3$ and is defined 
as
\begin{equation}
\label{defuniform}
\chi_3 = \frac{1}{L^2} \int_0^\beta dt \ \frac{1}{Z} \mbox{Tr}[M^3(0) M^3(t)
\exp(- \beta H)],
\end{equation} 
with $M^3 = \sum_x S^3_x$. Both $\chi_1$ and $\chi_3$ can be measured very 
efficiently with the loop-cluster algorithm using improved estimators 
\cite{Wie94}.

Another reference that provides analytic effective field theory results, which
can be compared with our Monte Carlo data, is a paper by Hasenfratz and
Niedermayer \cite{Has93}. Using magnon chiral perturbation theory up to
two-loop order, they obtained the finite-size and finite-temperature effects of 
$\chi_1$ in the $\epsilon$-regime
\begin{equation}
\label{chiscube}
\chi_1  = \frac{{\cal M}^2 L^2 \beta}{2} \left\{1 + \frac{c}{\rho L l} 
\beta_1(l) + \frac{1}{2} \left(\frac{c}{\rho L l}\right)^2 \left[\beta_1(l)^2 
+ \beta_2(l)\right] + {\cal O}\left(\frac{1}{L^3}\right) \right\}.
\end{equation}
The quantity $l = (\beta c /L )^{1/3}$ determines the shape of an approximately 
cubic  space-time box of volume $L \times L \times \beta$, with 
$\beta c \approx L$. The functions $\beta_i(l)$ are known shape-dependent 
coefficients \cite{Has90,Has93}. For an exactly cubical space-time volume 
(i.e.\ for $l = 1$) the result of eq.(\ref{chiscube}) agrees with 
eq.(\ref{moments}) since
\begin{equation}
\label{susc}
\langle (\Phi - \langle \Phi\rangle)^2\rangle + \langle \Phi\rangle^2 = 
\langle \Phi^2 \rangle = \frac{2 \chi_1 \widetilde{\cal M}^2}
{L^2 \beta {\cal M}^2} = \frac{2 \chi_1 a^4}{L^2 \beta}.
\end{equation}
The factor 2 arises due to the two components of the magnetization vector.
Remarkably, up to two-loop order the analogous expression for $\chi_3$ takes 
the simple form
\begin{equation}
\label{chiucube}
\chi_3 =  \frac{\rho}{c^2}+{\cal O}\left(\frac{1}{L^3}\right),
\end{equation}
which does not display any corrections of lower orders, neither of 
${\cal O}(1/L)$ nor ${\cal O}(1/L^2)$. 

The above expressions have been used to determine the low-energy parameters 
by a fit of $\chi_1$ and $\chi_3$ to Monte Carlo data \cite{Jia10.1}
\begin{equation} 
\label{parameters}
{\cal M} = 0.43561(1)/a^2,\quad \rho = 0.26974(5)J,\quad c = 1.1347(2)Ja.
\end{equation}
In this very accurate study, the cubical geometry has been reached by tuning 
$\beta$ until temporal and spatial winding numbers agreed. The spin-wave 
velocity has then been determined as $c = L/\beta$ with fraction of a permille
precision. The fitted magnetization density is consistent with the
result ${\cal M} = 0.437(2)/a^2$ obtained in \cite{San99}. For the 2-d 
spin $\tfrac{1}{2}$ Heisenberg model, using the same method, the corresponding 
low-energy parameters have recently also been determined with fraction of a 
permille accuracy in \cite{Jia10.2}.

\section{Probability Distribution of the Magnetization}

The quantum XY model can be simulated very efficiently with the loop-cluster algorithm \cite{Eve93,Wie94,Bea96}. In \cite{Har97} the Kosterlitz-Thouless phase transition has been studied for the first time using the loop-cluster algorithm. The transition temperature $T_{KT}$ has been determined very precisely from the winding numbers, which was not possible before. By using the same improved estimator as 
introduced in \cite{Ger09} we extract the probability distribution of the 
magnetization. Every cluster contributes additively to the first component of 
the magnetization. The cluster size $|{\cal C}|$ (i.e.\ the number of lattice 
points in a given cluster) determines the first component of the magnetization 
of the cluster ${\cal C}$, which is proportional to $\pm |{\cal C}|$. Under 
cluster flip the magnetization of a cluster changes sign. Starting from a given 
spin configuration, the distribution of the magnetization is recorded as a 
histogram which is built iteratively using one cluster after another. The 
initial histogram $p_1(m)$ (with $m \in \{-M,-M+1,...,0,...,M-1,M\}$, where $M$ 
is the number of space-time lattice points) is constructed from the first 
cluster as
\begin{equation}
p_1(m) = \frac{1}{2} \left[\delta_{m,|{\cal C}_1|} + 
\delta_{m,- |{\cal C}_1|}\right].
\end{equation}
The two entries of the initial histogram correspond to the two possible 
orientations of the first cluster, each arising with probability $\tfrac{1}{2}$.
In the $i$-th iteration step (with $i \in \{2,3,...,N\}$), where $N$ is the 
number of clusters of a given configuration, a new histogram $p_i(m)$ is built 
from the previous one as
\begin{equation}
p_i(m) = \frac{1}{2}\left[p_{i-1}(m + |{\cal C}_i|) +
p_{i-1}(m - |{\cal C}_i|)\right].
\end{equation}
The final histogram after $N$ steps is
given by $p_N(m)$. In figure 1 we show examples of histograms $p_N(m)$ obtained for two individual 
spin configurations. In the left panel the example contains one cluster that is bigger than all the other clusters together. 
Therefore the region around $m = 0$ is not sampled. Additionally, there are two
relatively large clusters that give rise to the multiple peaks in the
distribution. On the other hand, in the example shown in the right panel, there
are two clusters of similar size, such that the region around $m = 0$ is also 
sampled.
\begin{figure}
\begin{center}
\epsfig{file=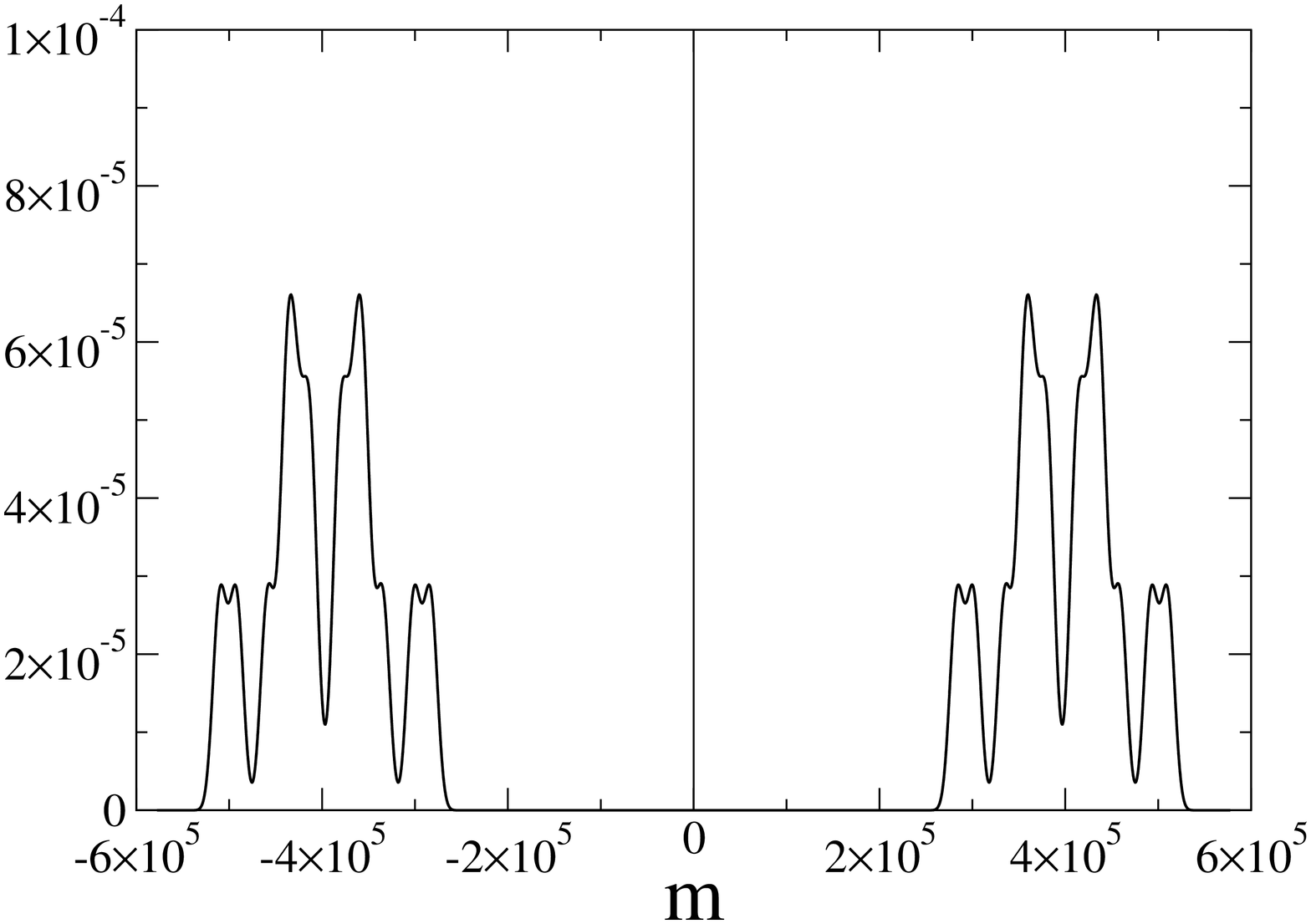,angle=-90,width=7.2cm}
\epsfig{file=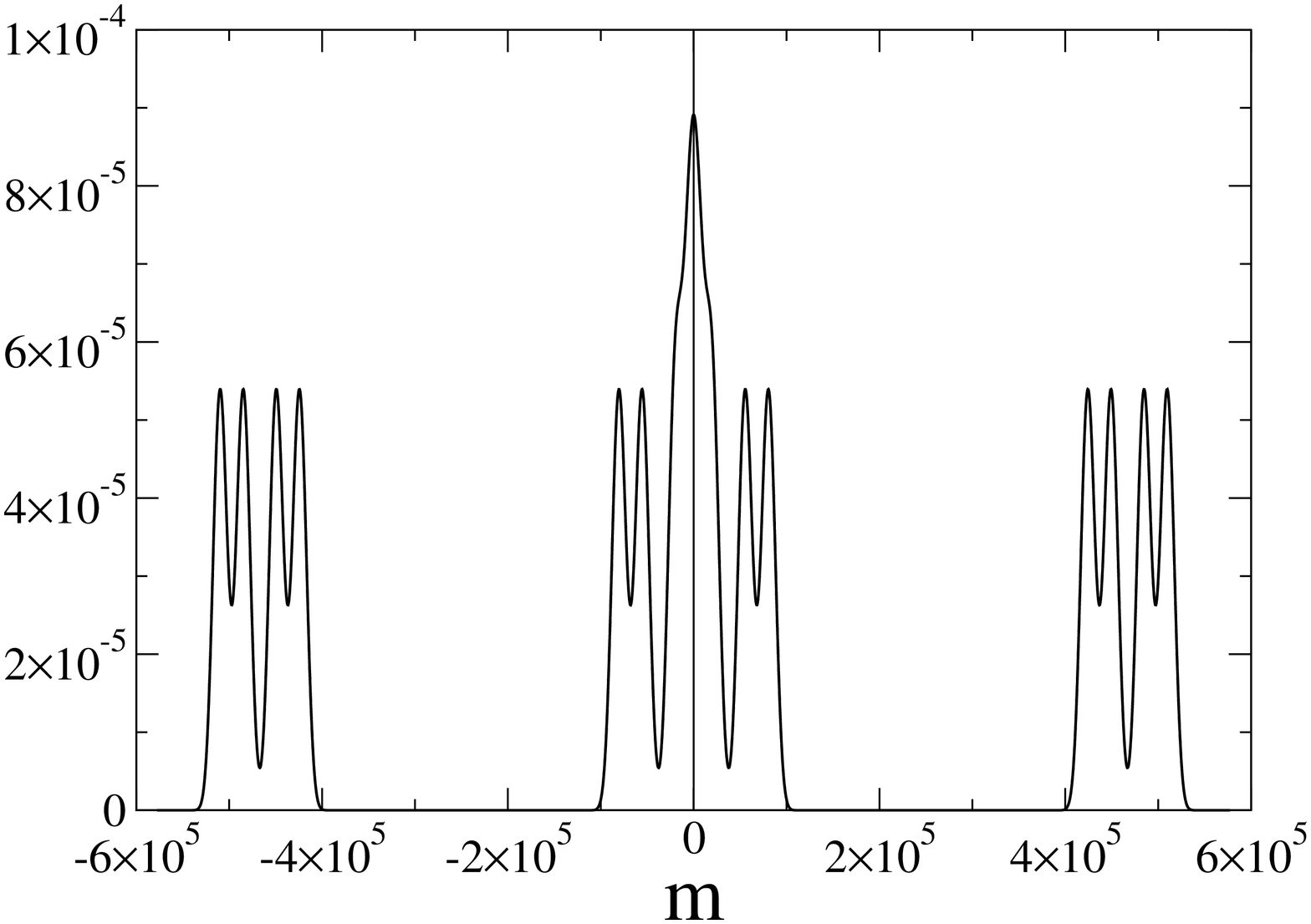,angle=-90,width=7.2cm}
\end{center}
\caption{\it Examples of histograms $p_{N}(m)$ obtained for two individual spin 
configurations on a $16^2$ lattice using the improved estimator.}
\end{figure}
The average of the histograms $p_N(m)$ for all configurations in the Markov 
chain generated by the cluster algorithm yields the final probability distribution of the first component of the magnetization 
\begin{equation}
p(m) = \langle p_N(m) \rangle.
\end{equation}
By construction, it is
properly normalized as
\begin{equation} 
\sum_{m = -M}^M p(m) = 1.
\end{equation}
The numerical effort to build the improved estimator is proportional to the 
number of lattice points $M$ and, in addition, proportional to the number of 
clusters $N$. Since the number of clusters is proportional to the volume, the 
evaluation of the improved estimator requires a computational effort 
proportional to $M^2$, and thus becomes rather time-consuming for large 
volumes. Still, the improved estimator increases the statistics by a factor of 
$2^N$ which is exponential in the volume. Therefore investing a polynomial 
effort $M^2$ should still be justified. Compared to \cite{Ger09} where the
method had been implemented in a straightforward manner, we have been able to
achieve substantial improvements. This allows us to investigate volumes as large
as $64^2$ while our original study of the Heisenberg model was limited to
$24^2$ lattices. 

Remarkably, the computational effort to build the histograms 
can be reduced by a factor of order up to 1000 (for a volume $64^2$) by several optimizations.
First of all, it is obvious that one should evaluate eq.(3.2) only for
$m \in \{ -M_i, -M_i+1, ..., 0, ..., M_i-1, M_i \}$
with
\begin{equation}
M_i = \sum _{j=1}  \left| {\cal C}_j \right|
\end{equation}
for the $i$-th iteration step. Therefore most of the zero values at the border of the partial histograms are not evaluated explicitly.
By sorting the clusters by their sizes and
building the histogram by starting with the smallest clusters and finishing
with the largest, $M_i$ and therefore the computational effort per iteration grows as slowly as possible.
Due to the time-discretization \cite{Wie94} (which has four Trotter steps), all 
cluster sizes $|{\cal C}_i|$ are multiples of~2.
Hence the cluster sizes can be divided by 2 when building the histograms.
This halves memory usage and it also halves the loop sizes for evaluating equation (3.2).
A further optimization is to use two arrays for the partial histograms.
One is the source and the other the destination for the new histogram.
The roles alternate per iteration.
This prevents some memory allocations, copy operations and initializations with zeros.
The histogram is symmetric. So only the part for $m \in \{0,...,M-1,M\}$ is held in memory and equations (3.1) and (3.2) are modified as
\begin{equation}
p_1(m) = \frac{1}{2} \delta_{m,|{\cal C}_1|},
\end{equation}
and
\begin{equation}
p_i(m) = \frac{1}{2}\left[p_{i-1}( \left| m + |{\cal C}_i| \right| ) +
p_{i-1}( \left| m - |{\cal C}_i| \right| )\right].
\end{equation}
Generating a cluster size histogram for one configuration one observes a relatively large number of clusters with the same small cluster size.
Hence, an improvement is to treat clusters of equal sizes in one iteration using the fact,
that a histogram of an even number of clusters of only one equal size equals to
\begin{equation}
p_{\text{even}}(m) = \frac{1}{2^{n_{|{\cal C}|}}}
 \sum _{k=0} ^{\frac{n_{|{\cal C}|}}{2}}
 \binom{n_{|{\cal C}|}}{\frac{n_{|{\cal C}|}}{2} +k}
 \delta_{\left| m \right|, 2 k |\cal C|},
\end{equation}
where $n_{|{\cal C}|}$ is the number of clusters of size $|{\cal C}|$.
For odd $n_{|{\cal C}|}$ the corresponding histogram equals to
\begin{equation}
p_{\text{odd}}(m) = \frac{1}{2^{n_{|{\cal C}|}}}
 \sum _{k=0} ^{\frac{n_{|{\cal C}|} -1}{2}}
 \binom{n_{|{\cal C}|}}{\frac{n_{|{\cal C}|} +1}{2} +k}
 \delta_{\left| m \right|, \left( 2 k +1 \right) |\cal C|}.
\end{equation}
A further improvement uses the fact that partial histograms built by eq.(3.1) always contain zero values either for odd or for even $m$. This holds for arbitrary cluster sizes $|{\cal C}_i|$.
By tracking these two cases and using only the non-zero values one can further optimize the computational effort.
All these optimizations do not influence the resulting histogram. For large volumes the computational effort is still growing with the square of the volume.

The final optimization is more delicate. We have found that it is possible to divide the cluster size by some volume-dependent factor larger than 2. The round off error is treated by an error propagation technique, which is currently not yet fully optimized. The division of the cluster size alters the resulting histogram. The dividing factor is chosen empirically, such that the resulting systematic error is smaller than the statistical error of the Monte Carlo data. For our simulations the dividing factor was proportional to $L$. The computational effort grew with a power of about 1.3 of the volume. This method could also be used for simulations in continuous time which result in non-integer cluster sizes.

The mean value of the first component of the magnetization $\Phi_1$ 
corresponding to a given value of $m$ is
\begin{equation}
\Phi_1 = \frac{m}{2 M}.
\end{equation}
The factor 2 arises because we are dealing with quantum spins $\frac{1}{2}$. 
Now one can identify the probability distribution of the first component of the 
mean magnetization as
\begin{equation}
\widetilde p(\Phi_1) \ d\Phi_1 = p(m).
\end{equation}
It turns out that the non-zero 
entries of the histogram $p(m)$ correspond to values of $m$ which are multiples 
of 4. In order to eliminate artifacts of the Trotter decomposition, we perform a
binning of the histograms $p(m)$ with four consecutive points in each bin. This 
implies that
\begin{equation}
d\Phi_1 = \frac{8}{M}.
\end{equation}
Altogether, we obtain
\begin{equation}
\widetilde p(\Phi_1) = \frac{p(m)}{d\Phi_1} = 
\frac{M}{8} p(m), \quad \Phi_1 = \frac{m}{2M},
\end{equation}
with $m$ constrained to be a multiple of 4. By construction, in the Euclidean 
time continuum limit the resulting probability distribution is normalized as
\begin{equation}
\label{normal1}
\int_{-\infty}^\infty d\Phi_1 \ \widetilde p(\Phi_1) = 1.
\end{equation}

Using the loop-cluster algorithm in its discrete-time variant 
\cite{Eve93,Wie94}, we have simulated the spin $\tfrac{1}{2}$ quantum XY model 
on a square lattice with $L/a$ between $8$ and $64$ at inverse temperatures 
$\beta = L/c$. We have worked at a sufficiently small lattice spacing in 
Euclidean time, such that the systematic discretization errors are negligible 
compared to the statistical errors. The probability distribution 
$\widetilde p(\Phi_1)$ of the first component $\Phi_1$ of the magnetization 
has been obtained using the improved estimator described above. A typical 
distribution is shown in figure 2.
\begin{figure}
\begin{center}
\epsfig{file=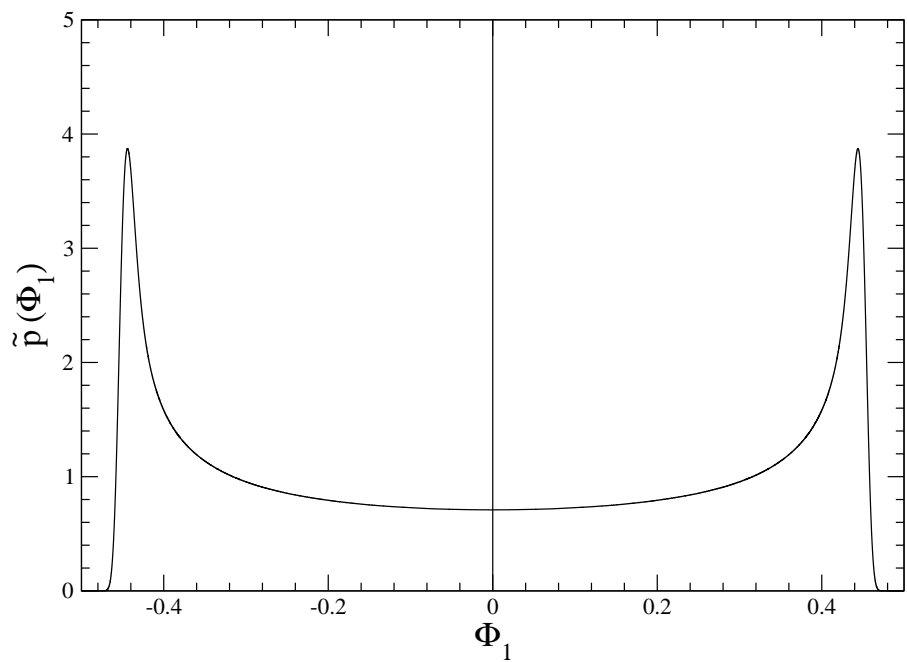,angle=-90,width=14cm}
\end{center}
\caption{\it Probability distribution $\widetilde p(\Phi_1)$ of the first 
component of the magnetization $\Phi_1$ on a $16^2$ lattice obtained with the 
improved estimator. The error bars of the distribution are of the order of the 
line width in this figure.}
\end{figure}
As we will see below, the information about the vicinity of the minimum of the 
constraint effective potential $u(\Phi)$ is contained in the region of $\Phi_1$
where $\widetilde p(\Phi_1)$ has its maxima.

Due to the $U(1)$ symmetry of the Hamiltonian, the probability distribution of 
the magnetization $\widetilde p(\Phi_1)$ depends only on the magnitude of the 
order parameter $\vec \Phi$. Consequently, the probability distribution 
$p(\Phi)$ can be expressed in terms of the probability distribution of the 
first component $\widetilde p(\Phi_1)$ by using the relation 
\begin{equation}
\widetilde p(\Phi_1) = \int_0^{2\pi} d\varphi \int_0^\infty  d\Phi  \ \Phi \
p(\Phi) \ \delta( \Phi_1 - \Phi \, \cos \varphi),
\end{equation}
which can be cast into the form
\begin{equation}
\widetilde p(\Phi_1) = 2 \int_{\Phi_1}^{\infty} d\Phi \
\frac{1}{\sqrt{1-{(\Phi_1/\Phi)}^2}} \ p(\Phi).
\end{equation}
The above relation is known as the Abel transform of the quantity $p(\Phi)$ and 
can be inverted, provided that both $p(\Phi)$ and its derivative $p'(\Phi)$ 
tend to zero faster than $1/\Phi$ as $\Phi \to \infty$. For the probability
distribution $p(\Phi)$ one then obtains 
\begin{equation}
\label{distributions}
p(\Phi) = - \frac{1}{\pi} \int_{\Phi}^{\infty} d\Phi_1
\frac{d \widetilde p(\Phi_1)}{d \Phi_1} \frac{1}{\sqrt{{\Phi_1}^2-{\Phi}^2}}.
\end{equation}
Hence, given the Monte Carlo data for $\widetilde p(\Phi_1)$, the probability
distribution of the magnitude of the magnetization can be extracted. With the
above equations one readily checks that the probability distribution
$p(\Phi)$ is properly normalized
\begin{equation}
2 \pi \int_0^{\infty} d\Phi \ \Phi \  p(\Phi) = 1,
\end{equation}
provided that $\widetilde p(\Phi_1)$ is normalized (see eq.(\ref{normal1})). We have determined the probability 
distributions $p(\Phi)$ from $\widetilde p(\Phi_1)$ using eq.(\ref{distributions}). In figure 3 some representative 
results for $2 \pi \Phi p(\Phi)$ are shown.
\begin{figure}
\begin{center}
\epsfig{file=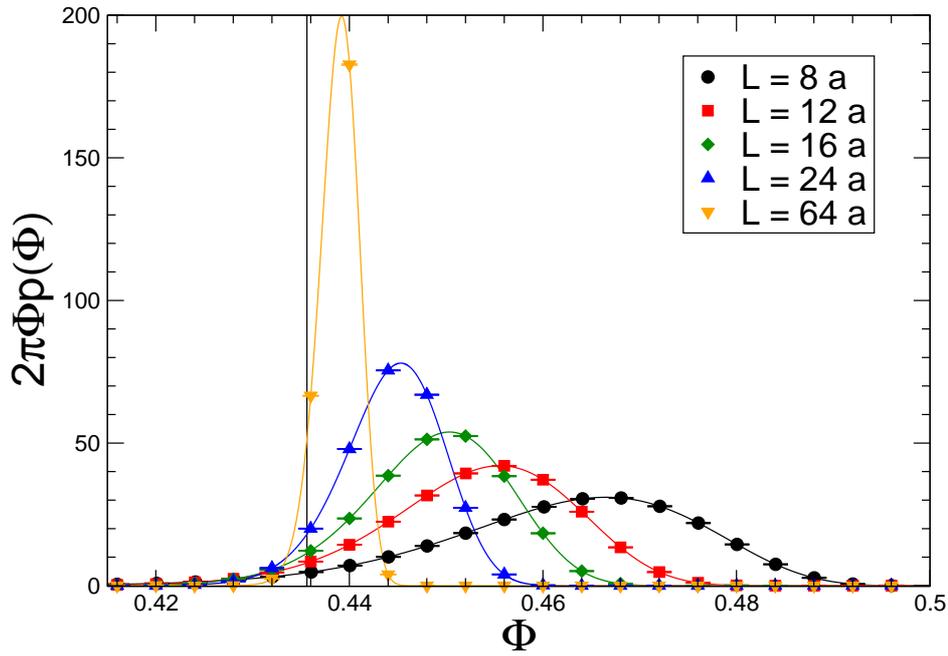,angle=-90,width=14cm}
\end{center}
\caption{\it Probability distributions $2 \pi \Phi p(\Phi)$ of the magnitude 
of the magnetization $\Phi$ for $L = 8a, 12a, 16a, 24a$, and $64a$. The lines are not fits but represent the Monte Carlo data themselves. The error bars are placed in equidistant positions of $\Phi$.
The vertical line at $\Phi = {\cal M} = 0.43561(1)$ represents the 
$\delta$-function distribution of the infinite system.}
\end{figure}
The mean value of $\Phi$ decreases as the volume increases, and the width of 
the distribution $p(\Phi)$ becomes narrower. It should be pointed out that the
distribution is not symmetric around its maximum. The distribution turns into a $\delta$-function in the infinite volume limit, centered at 
$\Phi = \widetilde{\cal M} = 0.43561(1)$. 

Furthermore, we compute the first and second centered moments $\langle \Phi\rangle$ and
$\langle (\Phi - \langle \Phi\rangle)^2 \rangle$ of the distribution $p(\Phi)$. 
Once we have computed $p(\Phi)$ this can be done in a straightforward manner using the 
improved estimator which requires a computational effort proportional to $M^{1.3}$.

\section{Comparison of Monte Carlo Simulations and Effective Theory Predictions}
In table 1 the first and second centered moments $\langle \Phi \rangle$ 
and $\langle (\Phi - \langle \Phi\rangle)^2\rangle$ of the distribution 
$p(\Phi)$ are compared with the effective field theory predictions of 
eq.(\ref{moments}). The errors of the theoretical predictions 
are due to the uncertainties in the low-energy parameters of 
eq.(\ref{parameters}) and due to neglecting higher-order corrections. For the first 
moment the agreement is very good for $L/a \geq 16$. The absolute value of the 
second moment is very small and its statistical error is relatively large. 
Still, there are systematic discrepancies between the Monte Carlo data of the first moment for small $L$, the second moment, and the ${\cal O}(1/L^2)$ effective theory predictions of eq.(\ref{moments}). This 
discrepancy is well accounted for by additional ${\cal O}(1/L^3)$ corrections. 
Such corrections involve next-to-leading low-energy parameters which multiply higher-order terms in the effective action. At order ${\cal O}(1/L^3)$ we would also have to evaluate 3-loop graphs. This
calculation, however, has not yet been worked out in the effective
theory. Parameterizing the 3-loop terms with unknown coefficients $\alpha_1$ and 
$\alpha_2$, i.e.\
\begin{eqnarray} 
&&\langle \Phi \rangle = 
\widetilde{\cal M} \left(1 + \frac{c}{\rho L} \frac{\beta_1}{2} +
\frac{c^2}{\rho^2 L^2} \frac{\beta_1^2}{8} \right) + 
\alpha_1 \left(\frac{c}{\rho L}\right)^3 + 
{\cal O}\left(\frac{1}{L^4}\right), 
\nonumber \\
&&\langle (\Phi - \langle \Phi\rangle)^2\rangle = 
\frac{\widetilde{\cal M}^2 c^2}{\rho^2 L^2} \frac{\beta_2}{2} + 
\alpha_2 \left(\frac{c}{\rho L}\right)^3 + 
{\cal O}\left(\frac{1}{L^4}\right), 
\end{eqnarray}
one obtains good fits to the Monte Carlo data for $\alpha_1 = - 0.0013(2)$ and
$\alpha_2 = - 0.00061(5)$. This shows that the Monte Carlo data are described 
well by the theoretical predictions. One may conclude that precise calculations of the two moments, combined with 3-loop predictions of the effective theory, would allow the determination of some combination of sub-leading low-energy parameters.
\begin{table}
\begin{center}
\begin{tabular}{|c|c|c|c|c|}
\hline
$L/a$ & $\langle \Phi \rangle_\text{MC}$ & $\langle \Phi \rangle_\text{theory}$
& $\langle (\Phi - \langle \Phi\rangle)^2\rangle_\text{MC}$ &
$\langle (\Phi - \langle \Phi\rangle)^2\rangle_\text{theory}$ \\
\hline
\hline
 8 & 0.46205(3) & 0.46224(1) & 1.90(7)$\times 10^{-4}$ & 2.7831(1)$\times 10^{-4}$ \\
\hline
 12 & 0.45305(10) & 0.45319(1) & 1.2(3)$\times 10^{-4}$ & 1.23694(6)$\times 10^{-4}$ \\
\hline
 16 & 0.44875(3) & 0.44873(1) & 4.6(8)$\times 10^{-5}$ & 6.958(4)$\times 10^{-5}$ \\
\hline
 20 & 0.44607(9) & 0.44608(1) & 4(2)$\times 10^{-5}$ & 4.453(2)$\times 10^{-5}$ \\
\hline
 24 & 0.44432(10) & 0.44432(1) & 7(27) $\times 10^{-6} $ * & 3.092(2)$\times 10^{-5}$ \\
\hline
 28 & 0.44307(9) & 0.44306(1) & 4(22) $\times 10^{-6} $ * & 2.272(1)$\times 10^{-5}$ \\
\hline
 32 & 0.44198(8) & 0.44212(1) & 4(2) $\times 10^{-5} $ * & 1.7394(9)$\times 10^{-5}$ \\
\hline
 40 & 0.44086(9) & 0.44081(1) & 9(25) $\times 10^{-6} $ * & 1.1132(6)$\times 10^{-5}$ \\
\hline
 48 & 0.43999(8) & 0.43994(1) & 6(200) $\times 10^{-7} $ * & 7.731(4)$\times 10^{-6}$ \\
\hline
 64 & 0.43880(9) & 0.43885(1) & 4(250) $\times 10^{-7} $ * & 4.3495(2)$\times 10^{-6}$ \\
\hline
\end{tabular}
\end{center}
\caption{\it Comparison of Monte Carlo data (MC) for the first and second 
centered moments $\langle \Phi \rangle$ and
$\langle (\Phi - \langle \Phi\rangle)^2\rangle$ of $p(\Phi)$ with predictions of 
the effective theory given by eq.(\ref{moments}). The numerical errors of the 
analytical expressions are due to small uncertainties in the values of the 
low-energy parameters quoted in eq.(\ref{parameters}). The entries with an asterisk (*) are statistically consistent with zero. The discrepancies between the Monte Carlo data and the effective field theory results are due to 
3-loop corrections that were neglected in the theoretical predictions.}
\end{table}

Starting from the probability distribution $p(\Phi)$ one obtains the constraint 
effective potential $u(\Phi)$ by using the relation 
$p(\Phi) = {\cal N} \exp(- L^3 u(\Phi))$. The constraint effective potentials 
corresponding to the curves in figure 3 are displayed in figure 4.
\begin{figure}
\begin{center}
\epsfig{file=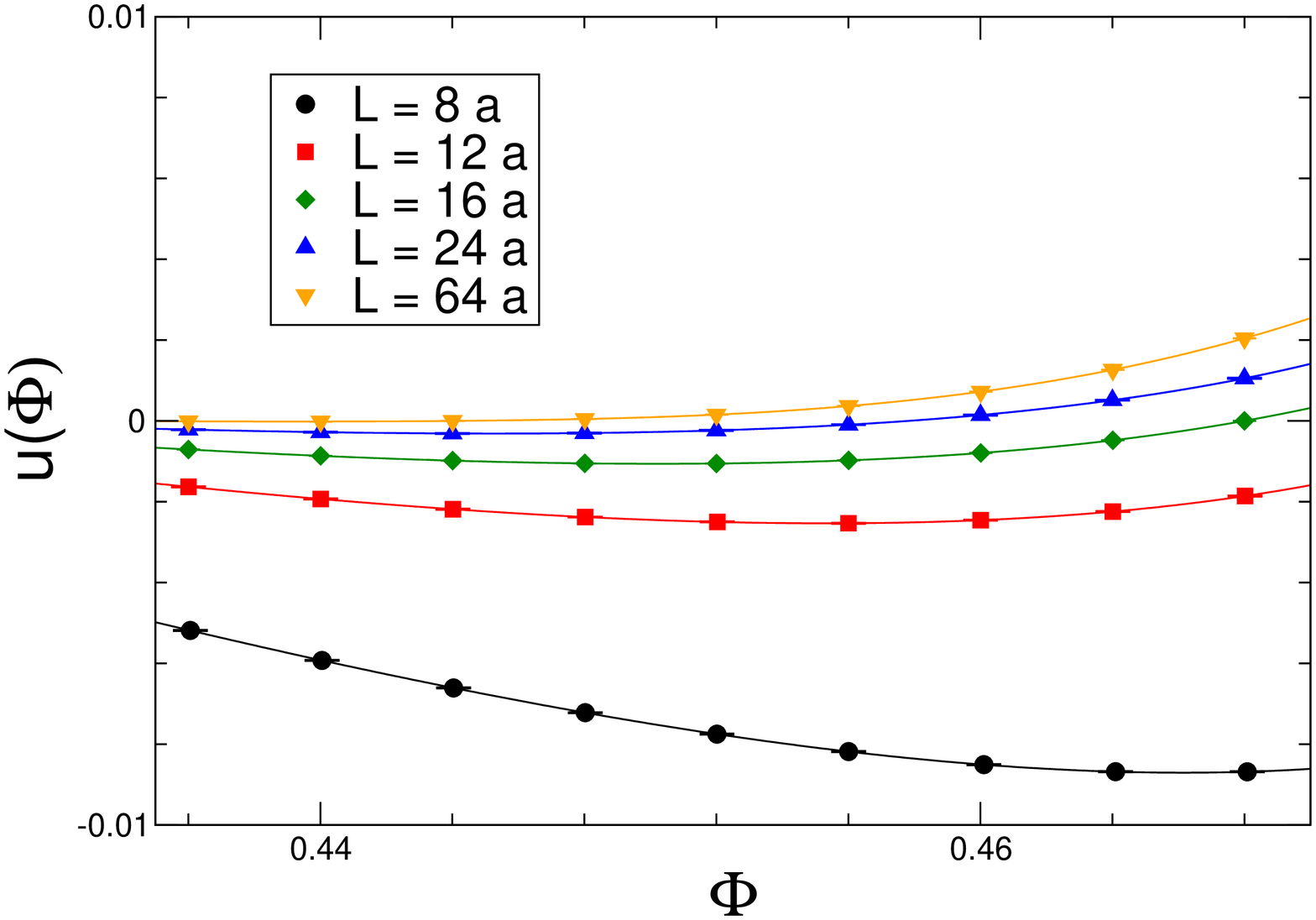,angle=-90,width=14cm}
\end{center}
\caption{\it Constraint effective potentials $u(\Phi)$ as functions of the 
magnitude of the magnetization $\Phi$ for $L = 8a, 12a, 16a, 24a$, and
$64a$. The constraint effective potential approaches a convex effective 
potential in the infinite volume limit.}
\end{figure}
With increasing volume the constraint effective potential approaches the 
effective potential, which is known to be a convex function. Using the rescaled 
variable $\psi = (\rho L/c) (\Phi - \widetilde{\cal M})/\widetilde{\cal M}$, 
one can also consider the extensive quantity $U(\psi) = L^3 u(\Phi)$ which is 
shown in figure 5.
\begin{figure}
\begin{center}
\epsfig{file=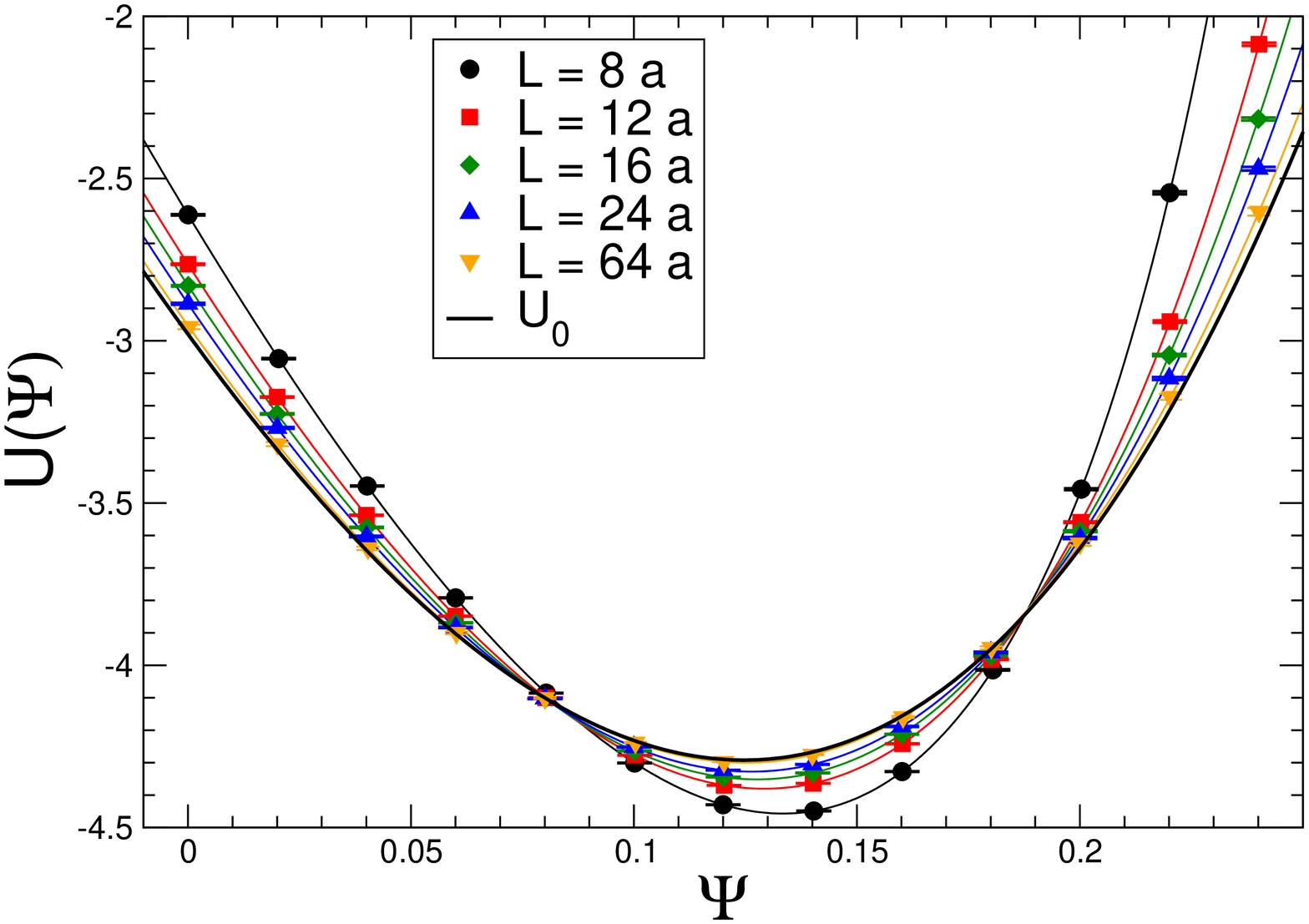,angle=-90,width=14cm}
\end{center}
\caption{\it The extensive quantity $U(\Phi)$ as a function of the rescaled 
variable $\psi = (\rho L/c) (\Phi - {\cal M})/{\cal M}$ for 
$L = 8a, 12a, 16a, 24a$, and $64a$, compared to the analytic infinite volume 
result $U_0(\psi)$.}
\end{figure}
Expanding $U(\psi) = U_0(\psi) + (c/\rho L) \ U_1(\psi) + {\cal O}(1/L^2)$, 
we have computed the universal part $U_0(\psi)$ by using Monte Carlo data for 
$L/a$ between $8$ and $64$. Some values of the function $U_0(\psi)$ extracted 
from the numerical data are compared with the analytic result of eq.(\ref{U0}) 
in figure 6. It should be pointed out that the observed agreement does not rely 
on any adjustable parameters. Even the normalization constant ${\cal N}$ of 
eq.(\ref{normalization}), which fixes an additive constant in the constraint 
effective potential, is predicted by the effective theory.
\begin{figure}
\vspace{1cm}
\begin{center}
\epsfig{file=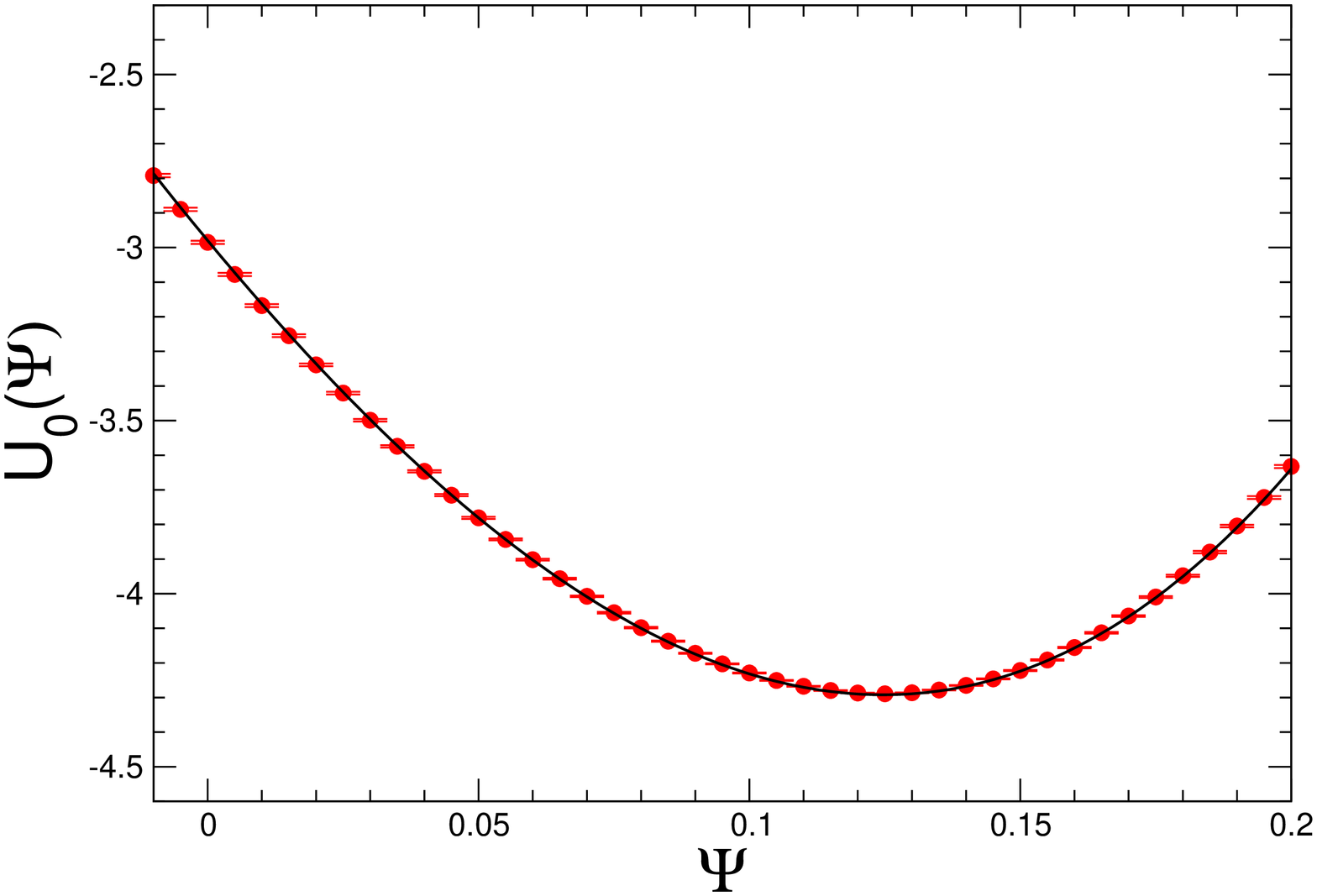,angle=-90,width=14cm}
\end{center}
\caption{\it The analytic result for the universal function $U_0(\psi)$ 
compared with numerical values obtained from the Monte Carlo data for
$U(\Phi)$. The comparison does not involve any adjustable parameters.} 
\end{figure}
As quantified in table 2, in the interval $\psi \in [0,0.2]$, i.e.\ around
the minimum of the constraint effective potential, the theoretical values of 
$U_0(\psi)$ and the numerical data agree remarkably well.
\begin{table}
\begin{center}
\begin{tabular}{|c|c|c|c|c|}
\hline
$\psi$ & $U_0(\psi)_\text{MC}$ & $U_0(\psi)_\text{theory}$ \\
\hline
\hline
0 & - 2.985(6)  & - 2.980 \\
\hline
0.05 & - 3.781(4) & - 3.781 \\
\hline
0.1 & - 4.229(2)  & - 4.232 \\
\hline
0.15 & - 4.222(2)  & - 4.224 \\
\hline
0.2 & - 3.632(5) & - 3.640 \\
\hline
\end{tabular}
\end{center}
\caption{\it Comparison of Monte Carlo data (MC) for the universal function
$U_0(\psi)$ with the effective theory prediction of eq.(\ref{U0}).}
\end{table}
To determine the low-energy constant $k_0$, we have used the extracted data for $U_1(\psi)$ as well as their theoretical prediction of eq.(\ref{U1}). A fit of the data to eq.(\ref{U1}) leads to $k_0=-0.0027(2)$. The result is illustrated in figure 7. Furthermore, $k_0$ can also be determined from 
fitting the $B$-dependent field expectation values $|\langle {\vec \Phi} \rangle (B)|$  
to their theoretical prediction of eq.(\ref{PhiB}). Since such an analysis provides 
a good check for the quantitative correctness of the numerical value for $k_0$ determined from 
$U_1(\psi)$, we calculate $|\langle {\vec \Phi} \rangle (B)|$ for several values of 
the magnetic field $B$ and the box size $L$ by using the technique of reweighting. 
Table 3 contains the results of $|\langle \vec \Phi \rangle(B)|$ obtained from 
reweighting. Using the data in table 3 as well as the corresponding theoretical 
prediction of eq.(\ref{PhiB}), we arrive at $k_0 = -0.0026(3)$ which is in excellent 
agreement with $k_0 = -0.0027(2)$ calculated from $U_1(\psi)$. The statistical consistency between $k_0 = -0.0026(3)$ determined from table 3 and the $k_0$-value obtained from $U_1(\psi)$ also demonstrates the reliability of the reweighting technique employed for the determination of the data in table 3. Indeed, we have observed consistency between the data in table 3 with the largest $B$-field for each $L$ and the corresponding $|\langle {\vec \Phi} \rangle (B)|$ determined by switching on explicitly a uniform magnetic field in the Monte Carlo simulations. Since for each $L$, the largest $B$ imposes the greatest challenge for the reweighting method, we conclude that all the data in table 3 obtained by reweighting are indeed quantitatively correct.

\begin{table}
\begin{center}
\begin{tabular}{|c|c|c|c|}
\hline
$L/a$ & $B/J$ & $|\langle \vec \Phi \rangle(B)|$ \\
\hline
\hline
64 & 0.00306135 & 0.44053(3) \\ 
\hline
64 & 0.0042 & 0.44142(4) \\ 
\hline
72 & 0.00328 & 0.44076(4) \\ 
\hline
72 & 0.00386 & 0.44121(7) \\ 
\hline
80 & 0.00266 & 0.44028(4) \\ 
\hline
80 & 0.00313 & 0.44069(5) \\ 
\hline
88 & 0.0022 & 0.43985(3) \\ 
\hline
88 & 0.00258 & 0.44021(6) \\ 
\hline
96 & 0.00184 & 0.43950(3) \\ 
\hline
96 & 0.00217 & 0.43984(5) \\ 
\hline
\end{tabular}
\end{center}
\caption{\it Monte Carlo data for $|\langle \vec \Phi \rangle(B)|$ which are
used in the determination of $k_0$.}
\end{table}

\begin{figure}
\begin{center}
\epsfig{file=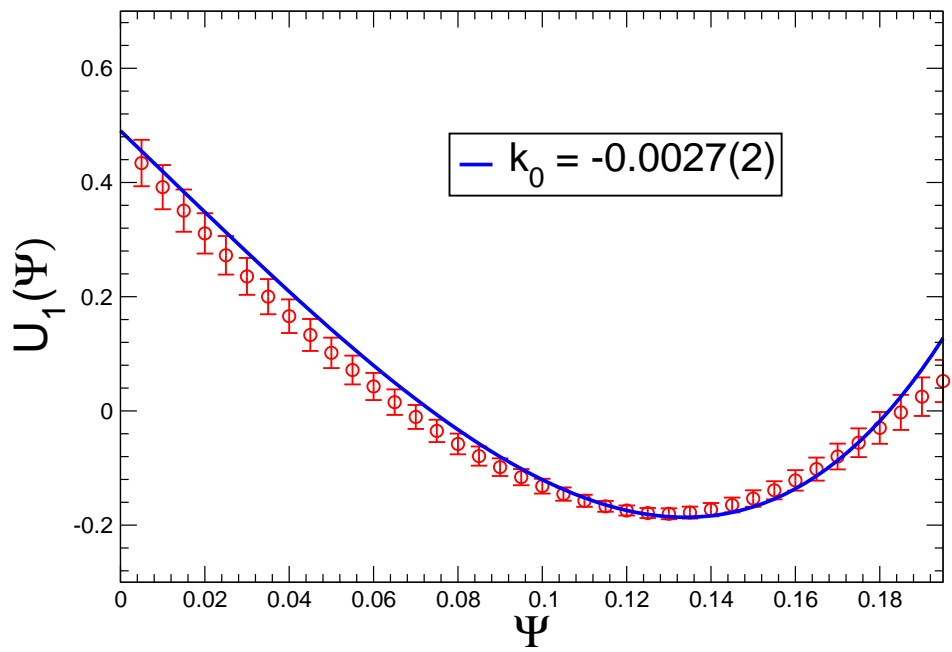,angle=-90,width=14cm}
\end{center}
\caption{\it Result of fitting the Monte Carlo data for $U_1(\psi)$ to the theoretical prediction.}
\end{figure}

\section{Conclusions}

We have computed the probability distribution of the magnetization in the $(2+1)$-d 
XY model by using an improved estimator as first proposed in \cite{Ger09}.
Substantial improvements in the implementation of the method have allowed us to
reach spatial volumes as large as $64^2$ (or even $96^2$ to calculate $|\langle {\vec \Phi} \rangle (B)|$). Using the improved estimator in a loop-cluster
algorithm simulation, we have determined the first and second moments 
$\langle \Phi \rangle$ and $\langle (\Phi - \langle \Phi\rangle)^2\rangle$ of 
the distribution $p(\Phi)$ of the magnitude $\Phi$ (of the mean magnetization 
vector $\vec \Phi$), as well as the constraint effective potential $u(\Phi)$ 
(obtained from $p(\Phi) = {\cal N} \exp[- L^3 u(\Phi)]$) for different 
space-time volumes. The Monte Carlo data are in excellent quantitative agreement
with analytic predictions which G\"ockeler and Leutwyler derived from a 
systematic low-energy effective field theory. This shows that the magnon 
effective field theory indeed provides a quantitatively correct systematic 
derivative expansion of the low-energy physics. Thanks to the very efficient 
loop-cluster algorithm, in the context of the $(2+1)$-d quantum XY model we 
were able to test theoretical predictions of the effective field theory up to 
two-loop order.

\section*{Acknowledgments}

We have benefited from correspondence and discussions with M.\ G\"ockeler, F.\ Niedermayer, and H.\ Leutwyler. C.\ P.\ H.\ , F.-J.\ J.\ and G.\ P.\ would like to thank the members of the Institute for Theoretical 
Physics at Bern University for their hospitality. U.\ G.\ would like to thank 
the members at the Departamento de F\'isica at Universidad de Santiago de Chile 
for their hospitality and the inspiring working atmosphere during a visit 
at which most of this manuscript was written. F.-J.\ J.\ is partially supported by NSC and NCTS (North). G.\ P.\ was partially supported by Dicyt 
grant 040931PA. The work of C.\ P.\ H.\ is supported by CONACYT Grant No.\ 
50744-F. This work is supported in parts by the Schweizerischer Nationalfonds 
(SNF). The ``Albert Einstein Center for Fundamental Physics'' at Bern University
is supported by the ``Innovations- und Kooperationsprojekt C-13'' of the 
Schweizerische Uni\-ver\-si\-t\"ats\-kon\-fe\-renz (SUK/CRUS).

\bibliography{biblio}

\end{document}